\documentstyle[11pt,twoside,newpasp,epsf,natbib]{article}
\def\edcomment#1{\iffalse\marginpar{\raggedright\sl#1\/}\else\relax\fi}
\reversemarginpar

\newcommand{\mem}[1]{\mathrm{ #1}}

\newcommand{\msun}{\, {\rm M}_\odot}

\newcommand{\spr}{\mbox{$s$-process}}

\begin{document}
\title{Fe-deficiency in H-deficient post-AGB stars due to n-capture nucleosynthesis
}
 \author{Falk Herwig}
\affil{University of Victoria, Victoria, BC, Canada}
\author{Maria Lugaro}
\affil{IoA, University of Cambridge, Cambridge, UK}
\author{Klaus Werner}
\affil{Universit\"at T\"ubingen, T\"ubingen, Germany}

\begin{abstract}
H-deficient  post-AGB  objects, e.g.\ PG1159  type  star  K1-16  and  
born-again  
AGB  star  Sakurai's object,  have been reported to be significantly
iron-deficient.  We find that the iron deficiencies expected due
to neutron-capture nucleosynthesis during either the progenitor AGB
evolution and/or the neutron burst that occurs as a result of the
rapid burning of protons during a post-AGB He-flash are generally in line
with  observations.  
\end{abstract}

Recently, Miksa et al.\ (2001) reported that the PG1159 central star (CS) K1-16
is deficient in iron 
by at least a factor 10-100 (see also contributions by Werner, 
Gr\"afener et al.\ and Crowther at this conference and Miksa et al.,\ 2002).  
PG1159 stars are likely the result of a
born-again evolution associated 
with a thermal pulse (TP) occurring during the   
post-Asymptotic Giant Branch phase (post-AGB). In addition, 
Sakurai's object, which experienced a very late TP 
(VLTP, see Herwig et al.\ (1999), 
shows an Fe-depletion of about 1dex and a
Fe/Ni ratio of only $\sim 3$ (Asplund et al., 1999).

The material of H-deficient post-AGB stars reflects the intershell
abundance of the progenitor AGB star with some superimposed
modification due to the mixing and/or nucleosynthesis processes which
lead to the formation of the H-deficiency. 
It has been established that the \spr\ in  AGB stars
is the result of recurrent neutron exposure episodes in the 
region between the He- and H-burning shell.
Neutron exposures lead to a significant depletion of iron
in this layer.
Therefore the final AGB intershell abundances have to be taken as initial
conditions for VLTP nucleosynthesis models.
These have been obtained with the AGB $s$-process code
TOSP (Gallino et al., 1998), which computes the recurrent neutron exposures
\begin{figure}
\plottwo{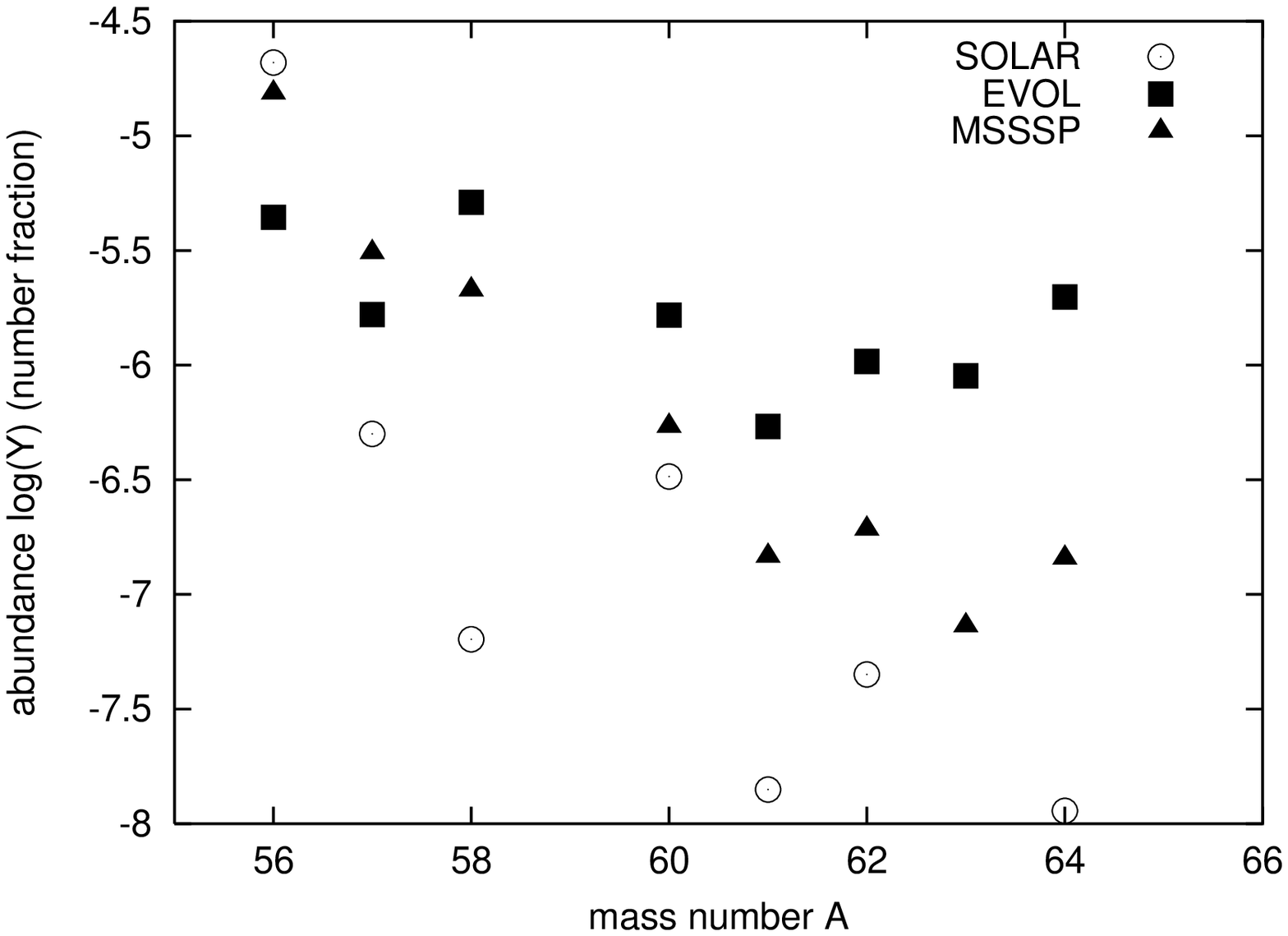}{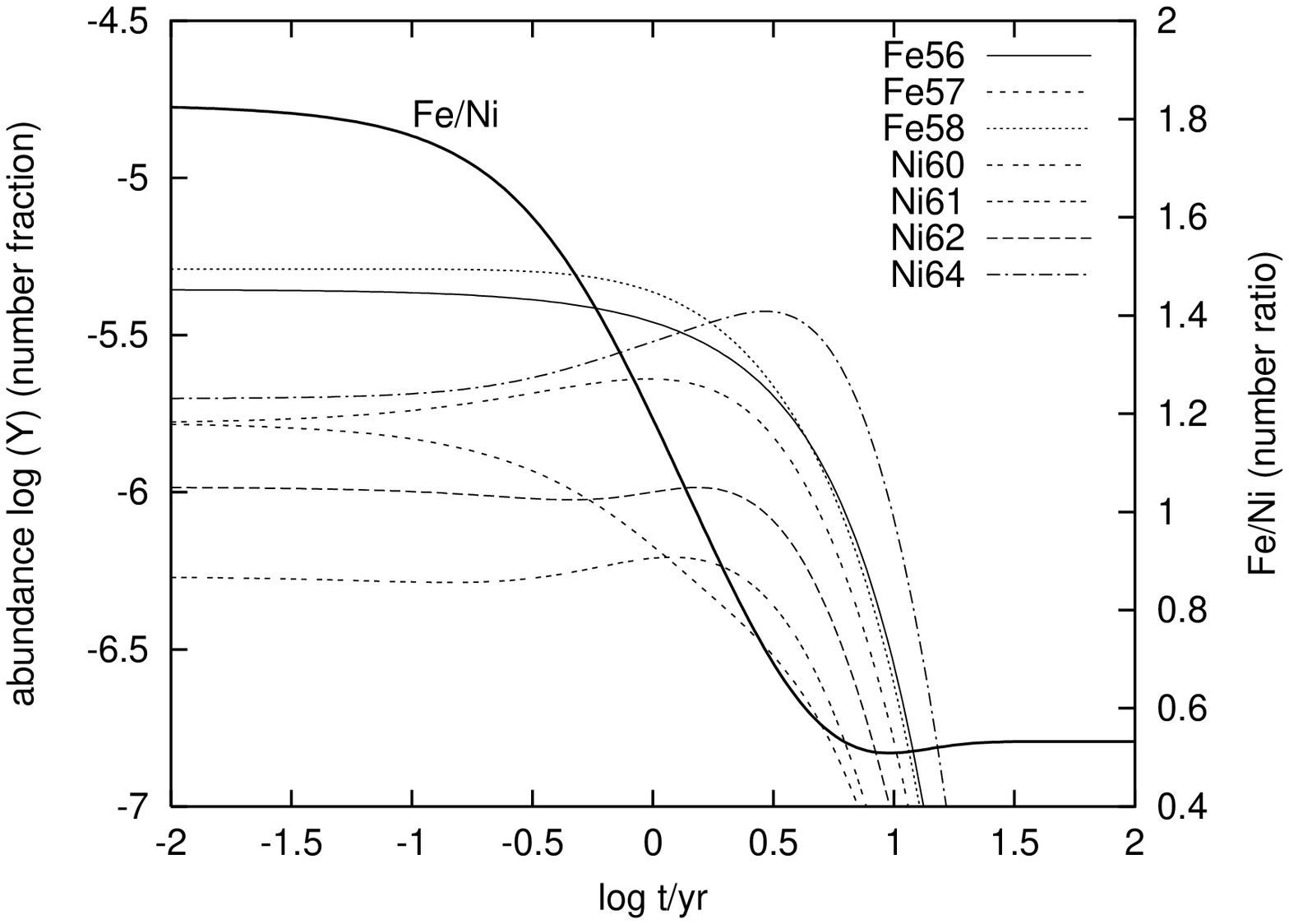}   %~/L/NW-ABgrains/T90.125X0.01.newdxnw05.mariaIS  %~/L/NW-ABgrains/T90.125X0.01.newdxnw05.mariaIS
\caption{
\textbf{Left:} Fe (A=56-58) and Ni (A=60-64) isotopic abundances in
the intershell at the end of the 
AGB evolution ($3\msun$, Z=0.02) as computed with the TOSP
post-processing code, using either the stellar evolution sequences of
the \textsf{EVOL} code with overshoot or the \textsf{MSSSP} code without
overshoot as input (for more details see Lugaro \& Herwig, 2001) and
for comparison the \textsf{SOLAR} abundances.  \textbf{Right:}
Fe and Ni isotopes from a one-zone model which simulates a VLTP  
($T_{\mem{8}}=1.25$, $X_{\mem{ini}}(\mem{H})=0.01$). 
Initial abundances are shown left, case \textsf{EVOL}.}
\end{figure}
and mixing events over many TPs using thermodynamic data from
full stellar evolution calculations. As shown in Fig.\,1
(left panel) Fe is noticeably depleted in the intershell of evolved AGB
stars and the results depend significantly on the stellar models on which
the \spr\ calculation is based.
We have created a number of one-zone nucleosynthesis models for
conditions  encountered during a VLTP (Herwig, 2001) 
with a fully implicit iterative nuclear network code.
The characteristic ingestion of hydrogen into the He-shell during a VLTP 
is  simulated by 
setting the initial H-abundance $X_{\mem{ini}}$ to a non-zero value.
We chose temperatures and densities as typically encountered at the
bottom of the H-ingestion convection zone. These conditions do persist
only for a fraction of a year. 
For the specific parameters of the simulation shown in Fig.\,1 (right
panel) a considerable depletion of iron in addition to the Fe-depletion from the 
AGB phase is
found. However, at this stage of modeling the exact numbers should not be
taken too seriously. The one-zone models
can only reveal the main processes involved and more elaborate models
will be constructed.\\
\noindent
\textbf{Acknowledgments:} F.H. would like to thank D.A. VandenBerg for support through his Operating Grant from the Natural Science and Engineering Research Council of Canada.

\end{document}